\documentclass[3p,times,twocolumn]{elsarticle}

\usepackage{ecrc}

\volume{00}

\firstpage{1}

\journalname{Nuclear Physics B Proceedings Supplement}

\runauth{}

\jid{nuphbp}

\jnltitlelogo{Nuclear Physics B Proceedings Supplement}

\usepackage{amssymb}

\begin{document}

\begin{frontmatter}




\title{Dispersive analysis of $\tau^-\to\pi^-\pi^0\nu_\tau$ Belle data}


\author[1]{Daniel G\'omez Dumm}
\author[2]{Pablo Roig}

\address[1]{IFLP, CONICET $-$ Dpto. de F\'{\i}sica,
Universidad Nacional de La Plata, C.C.\ 67, 1900 La Plata, Argentina.}
\address[2]{Grup de F\'{\i}sica Te\`orica, Institut de F\'{\i}sica d'Altes Energies,
Universitat Aut\`onoma de Barcelona, E-08193 Bellaterra, Barcelona, Spain.}

\begin{abstract}
We analyse Belle data on the decay $\tau^- \to \pi^- \pi^0 \nu_\tau$ using a
dispersive representation of the vector form factor which is consistent with
chiral symmetry and preserves analyticity and unitarity exactly. We fit the
unknown theoretical parameters from the data, determining the values of the
related low-energy observables $\left\langle r^2\right\rangle^\pi_V$ and
$c_V^\pi$. The implementation of isospin breaking effects is also
discussed~\cite{DY}.
\end{abstract}


\end{frontmatter}


\section{Introduction}
\label{Intro} The vector form factor of the pion, $F_V^\pi(s)$, encodes all
unknown strong interaction dynamics in $\tau^- \to \pi^- \pi^0 \nu_\tau$
decays~\cite{Cirigliano:2001er}. It is defined as
\begin{equation}
 <\pi^-(p_{\pi^-})\pi^0(p_{\pi^0})|\bar{d}\gamma^\mu
u|0>=\sqrt{2}(p_{\pi^-}-p_{\pi^0})^\mu F_V^\pi(s)\,,
\end{equation}
with $s=(p_{\pi^-}+p_{\pi^0})^2$. This form factor is not only relevant for the
understanding of the hadronization of QCD currents at low energies
(see~\cite{Talks1}), but also represents a crucial ingredient for the
evaluation of the leading order (LO) hadronic contribution to the anomalous
magnetic moment of the muon, which provides a very stringent probe of new
physics~\cite{Actis:2010gg}. In addition, from the high-energy perspective,
the $\pi\pi$ channel is (together with the three pion channel) essential to
follow the spin in the Higgs di-tau decay channels at LHC~\cite{Talks2}, and
thus to determine its spin and CP properties with the help of
TAUOLA~\cite{Talks3}.

\section{Theoretical setting}
In $\tau^- \to \pi^- \pi^0 \nu_\tau$ decays the electroweak part of the
process is theoretically under control, while the hadronization of the quark
currents is more involved since in the spanned energy region QCD is
essentially non perturbative. The approximate chiral symmetry of
light-flavoured QCD allows to build an effective field theory, known as
Chiral Perturbation Theory ($\chi$PT) \cite{Weinberg:1978kz, Gasser:1983yg,
Gasser:1984gg, Bijnens:1999sh}, which is able to describe accurately the
low-energy part of the spectrum, but fails at larger invariant masses
\cite{Colangelo:1996hs}. In this region, new particles ($\rho$, $K^\star$,
$a_1$, ...) are excited and their momenta and masses are large enough to
prevent their use in the expansion parameters of the effective theory.
Therefore these new active degrees of freedom have to be included in the
action, and a new expansion parameter is needed. In this respect, the
inverse of the number of colours of the QCD gauge group has proven to be a
useful quantity to build the expansion upon \cite{'tHooft:1973jz,
'tHooft:1974hx, Witten:1979kh}. A modelization of this idea in the meson
sector for three flavours is provided by Resonance Chiral Theory (R$\chi$T)
\cite{Ecker:1988te, Ecker:1989yg, Cirigliano:2006hb, Kampf:2011ty}, which
recovers the $\chi$PT results at next-to-leading order (NLO). Since it is
known that the lightest resonances dominate the dynamics, the infinite tower
of states predicted in the $N_C\to\infty$ limit can be restricted to the
first excitations, taking into account as many states as required by the
data. In addition, it is seen that the inclusion of resonance widths is
essential to describe the observed phenomenology, although widths arise at
NLO in the $1/N_C$ expansion. In principle, one can take into account this
effect by computing off-shell widths consistently within R$\chi$T
\cite{GomezDumm:2000fz, Dumm:2009va}. On the other hand, as in any effective
theory, the symmetry properties determine the operators allowed in the
Lagrangian, but leave the corresponding coupling constants unknown.
However, the QCD short-distance behaviour of the Green functions and related
form factors \cite{Ecker:1988te, Ecker:1989yg, Cirigliano:2006hb, Kampf:2011ty,
RuizFemenia:2003hm, Cirigliano:2004ue, Cirigliano:2005xn, Dumm:2009kj, Guo:2010dv, Dumm:2012vb} provide a set of
relations among these coefficients that renders R$\chi$T more predictive.

\section{Vector form factor of the pion and fits to data}
Different approaches have been developed to deal with the diverse energy
regimes. For $s<M_\rho^2$, the computation of $F_V^\pi(s)$ at NNLO in
$\chi$PT \cite{Gasser:1984ux, Gasser:1990bv, Bijnens:1998fm, Bijnens:2002hp} proves useful.
In order to enlarge the domain of applicability up to 1 GeV, unitarization
techniques \cite{DeTroconiz:2001wt, Oller:2000ug} and the Omn\`es solution
to the dispersion relation have been employed \cite{Guerrero:1997ku,
Pich:2001pj, Hanhart:2012wi}. Finally, in order to reach energies up to
$M_\tau$, the inclusion of the $\rho^\prime$ resonance
\cite{SanzCillero:2002bs} and even a tower of resonances, inspired in the
$N_C\to\infty$ limit \cite{Dominguez:2001zu, Bruch:2004py}, have been
proposed. From the R$\chi$T Lagrangian, including only the $\rho(770)$
multiplet one obtains
\begin{equation}
 F_V^\pi(s)=1+\frac{F_V G_V}{F^2}\frac{s}{M_V^2-s}\ ,
\label{ff0}
\end{equation}
where $F$ is the pion decay constant in the chiral limit, $M_V=M_\rho$, and
$F_V$ and $G_V$ measure the strength of the $\rho\pi\pi$ and $\rho V_\mu$
couplings, respectively, and $V_\mu$ stands for the quark vector current.
If the vanishing of the form factor at large energies is required, the
relation $F_V G_V = F^2$ is obtained, yielding
$F_V^\pi(s)=\frac{M_V^2}{M_V^2-s}$. Now one can do
better~\cite{Guerrero:1997ku} and match this expression to the NLO result in
$\chi$PT. Final state interactions are included through the so-called chiral
loop functions $A_P(s,\mu^2=M_\rho^2)$. Then, the unitarity and analyticity
constraints determine the Omn\`es exponentiation of the full loop function,
leading to
\begin{equation}
F_V^\pi(s)=\frac{M_V^2}{M_V^2-s}\mathrm{exp}\left\lbrace\frac{-s}{96\pi^2F^2}
\left[A_\pi(s)+\frac{1}{2}A_K(s)\right]\right\rbrace\,.
\end{equation}
Here one cannot simply include the resonance width by replacing $M_V^2-s$ by
$M_V^2-s-iM_V\Gamma_V(s)$ in the propagator, since this would double count
$\Im m[A_P(s)]$ and analyticity would be violated at NNLO in the chiral
expansion. We follow instead a procedure similar to that proposed
in Ref.~\cite{Boito:2008fq} for the $K\pi$ vector form factor, in which
unitarity and analyticity are satisfied exactly. The starting point is a
form factor as in Eq.~(\ref{ff0}), where the loop functions are resummed
into the denominator:
\begin{equation}
 F_V^\pi(s)^{(0)} = \frac{M_V^2}{M_V^2\left\lbrace1+\frac{s}{96\pi^2F^2}
 \left[A_\pi(s)+\frac{1}{2}A_K(s)\right]\right\rbrace-s}
\end{equation}
\begin{equation}
 = \frac{M_V^2}{M_V^2\left\lbrace1+\frac{s}{96\pi^2F^2}
\Re e\left[A_\pi(s)+\frac{1}{2}A_K(s)\right]
\right\rbrace-s-i M_V \Gamma_V(s)}\,.
\end{equation}
Thus the relevant ($I=1$, $J=1$) phase shift is taken to be
\begin{equation}
\mathrm{tan} \delta_1^1(s)=\frac{\Im m F_V^\pi(s)^{(0)}}{\Re e
F_V^\pi(s)^{(0)}}\,.
\end{equation}
This is now used as input for a three-subtracted dispersion relation for
the form factor. In this way one gets
\begin{eqnarray}\label{finalFF}
F_V^\pi(s) & = & \mathrm{exp}\left\lbrace \alpha_1 s+\frac{\alpha_2}{2}
s^2 \right.\nonumber \\
& & \left. +\frac{s^3}{\pi}\int_{4m_\pi^2}^\infty \mathrm{d}s^\prime
\frac{\delta_1^1(s^\prime)}{(s^\prime)^3(s^\prime-s-i\epsilon)}
\right\rbrace\,.
\end{eqnarray}

Kinematical isospin corrections can be easily included in
Eq.~(\ref{finalFF}) by considering different masses for the charged and
neutral pions and kaons. In addition, at the same order one should also take
into account electromagnetic corrections \cite{Cirigliano:2001er,
Cirigliano:2002pv}, which enter through a local term $f^{\rm elm}_{\rm
local}$ and a global factor $G_{EM}(s)$ \cite{FloresBaez:2006gf}. 
Thus we consider three possible expressions for the form factor to perform
our fits to Belle data:
\begin{itemize}
 \item Fit I, corresponding to $F_V^\pi(s)$ from Eq.~(\ref{finalFF}).
 \item Fit II, same as I but with the inclusion of kinematical corrections
 ($m_{\pi^\pm}\neq m_{\pi^0}$, $m_{K^\pm}\neq m_{K^0}$).
 \item Fit III, including kinematical and electromagnetic corrections.
\end{itemize}
Our fitting parameters are $M_\rho$, $F$, $\alpha_1$ and $\alpha_2$. It is
found that without the inclusion of additional resonances, one can obtain
good fits to Belle data for $s\lesssim 1.5$ GeV$^2$. The best fit results to
the first 30 points (central value of the bin corresponding to $1.525$
GeV$^2$, with $0.05$ GeV$^2$ bin width) are shown in Table \ref{Tab:1},
where we have considered the $1/N_{\pi\pi}dN_{\pi\pi}/ds$ distribution
measured by Belle (this includes error correlations). These fits show,
firstly, that the dispersive description of the form factor can indeed
successfully account for the experimental data up to $s\lesssim 1.5$
GeV$^2$, and secondly, that the approach employed by the Belle Collaboration
(named here as II) is indeed an adequate one, as it yields the lowest
$\chi^2/ndf$ values according to our fits. Notice that, given the low energy
threshold for this decay, the subtraction constants are fixed at a
relatively low energy scale, and the dispersive representation turns out to
be insensitive to the dynamics at large energies. In order to get a result
for the form factor that can be valid up to $s=M_\tau^2$, the expression in
Eq.~(\ref{finalFF}) can be e.g.~matched at intermediate energies to a
phenomenologically adequate function \cite{DY} like that given in
Ref.~\cite{Roig:2011iv} (included in the new version of TAUOLA
\cite{Shekhovtsova:2012ra}), or the Gounaris-Sakurai parametrization
\cite{Gounaris:1968mw} used by Belle \cite{Fujikawa:2008ma}.

\begin{table*}[h!]
 \begin{center}
\begin{tabular}{|c||c|c|c|}
\hline
Parameter & Fit Value(I) & Fit Value(II) & Fit Value(III)\\
\hline
$M_\rho$& $0.8431(5)((17))$ & $0.8280(4)((14))$ & $0.8276(4)((21))$\\
$F_\pi$ & $0.0901(2)((5))$& $0.0902(2)((4))$& $0.0906(2)((4))$\\
$\alpha_1$& $1.87(1)((3))$ & $1.84(1)((3))$ & $1.81(1)((2))$\\
$\alpha_2$& $4.29(1)((7))$ & $4.34(1)((7))$ & $4.40(1)((6))$\\
\hline
$\chi^2/ndf$ & $1.37$ & $1.39$ & $1.56$\\
\hline
$\Gamma_{\rho}(M_\rho^2$)& $0.207(1)((3))$ & $0.194(1)((3))$ & $0.192(1)((4))$\\
\hline
\end{tabular}
\caption{\small{Fit results to the Belle $1/N_{\pi\pi}dN_{\pi\pi}/ds$
distribution, including correlations between errors. The errors in single
and double brackets correspond to those arising only from the fit and those
obtained after considering theoretical systematics, respectively. Energy
units are given in GeV powers. $\Gamma_{\rho}(M_\rho^2$) is obtained using
the fitted values of $M_\rho$ and $F_\pi$ and is given only for reference.}}\label{Tab:1}
\end{center}
\end{table*}

In the table, the errors quoted in single brackets are those resulting only
from the fit, i.e.~neglecting the systematic errors arising from our
theoretical approach. These are e.g.~given by the energy range to be fitted,
the number of subtractions and the value of the upper integration limit in
the dispersive integral. In order to estimate the associated total errors we have
extended our fit up to energies in the range $[1.325,1.525]$ GeV$^2$, we
have taken into account the results for 2 and 4 subtractions, and we have
taken $s_\infty$ in the range $[4,\infty]$ GeV$^2$. In this way we end up
with the numbers quoted in double brackets in Table \ref{Tab:1}. Notice that
the input values for the $\rho$ mass and width still need to be translated
to the physical pole values, which are reasonably lower \cite{DY}.

It is seen that our results show a lower $\chi^2/ndf$ for our fitting
options I and II, compared to that obtained for option III. Thus, the best
agreement with the data is reached by including SU(2) isospin breaking
only kinematically, although comparable results are obtained in the isospin
symmetric case.

Finally, we have also computed the low-energy observables $\left\langle
r^2\right\rangle^\pi_V$ and $c_V^\pi$ appearing in the low-$s$ expansion of
$F_V^\pi(s)$. Our results are quoted in Table \ref{Tab:2}, together with
those obtained in previous analyses. It is seen that the values are entirely
compatible, while the errors are found to be slightly reduced thanks to the
good quality of present Belle data.

\begin{table*}[h!]
 \begin{center}
\begin{tabular}{|c|c|c|}
\hline
Determination & \small{$\left\langle r^2\right\rangle^\pi_V$ (GeV$^{-2}$)}& \small{$c_V^\pi$ (GeV$^{-4}$)}\\
\hline
Our fit & $10.86(14)$ & $3.84(3)$\\
Bijnens and Talavera \cite{Bijnens:2002hp} & $11.22(41)$ & $3.85(60)$\\
Pich and Portol\'es \cite{Pich:2001pj} & $11.04(30)$ & $3.79(4)$\\
\hline
\end{tabular}
\caption{\small{Low-energy observables of the vector pion form factor up to
the quadratic term. We give the results from our fit, the $\mathcal{O}(p^6)$ $\chi$PT
analysis in Ref.~\cite{Bijnens:2002hp} and the dispersive
analysis in Ref.~\cite{Pich:2001pj}.}} \label{Tab:2}
\end{center}
\end{table*}

\section{Conclusions}
We have elaborated a dispersive representation of $F_V^\pi(s)$, which
preserves analyticity and unitarity exactly and reproduces the low-energy
limit of $\chi$PT up to NLO. We have performed different fits to Belle
experimental data, which allow to determine our four input model parameters.
The fits show a good agreement with the data, and no significant improvement
is found after the inclusion of isospin breaking corrections. In addition,
from our fits we have evaluated the low-energy quantities $\left\langle
r^2\right\rangle^\pi_V$ and $c_V^\pi$, which turn out to be consistent with
previous determinations. Our framework is also able to provide good quality
fits to $\sigma(e^+e^-\to\pi^+\pi^-)$ scattering at low energies, which can
be used to determine $a_\mu^{\pi\pi,{\rm LO}}$ from $\tau$ decays and
$e^+e^-$ scattering consistently.




\section*{Acknowledgements}
P.R. congratulates the organizers for the profitable and enjoyable TAU'12
Conference where discussions on this topic with M. Davier, H. Hayashii, G. L\'opez Castro and 
A. Pich are acknowledged. We thank Hayashii Hisaki for providing us with Belle correlation
matrix. This work has been partially funded by CONICET (Argentina) under
grant \# PIP 02495, and by ANPCyT (Argentina) under grant \# PICT 2011-0113 and by the Spanish grants 
FPA2007-60323, FPA2011-25948 and by the Spanish Consolider Ingenio 2010 Programme CPAN (CSD2007-00042).
\nocite{*}
\bibliographystyle{elsarticle-num}
\bibliography{2pi}







\end{document}